\documentclass[%
 aip,
 amsmath,amssymb,
preprint,%
]{revtex4-1}

\usepackage{graphicx}
\usepackage{dcolumn}
\usepackage{bm}

\usepackage[utf8]{inputenc}
\usepackage[T1]{fontenc}
\usepackage{mathptmx}

\begin{document}

\preprint{AIP/123-QED}

\title[Adiabatic Afterglow]{Collisionless Adiabatic Afterglow}

\author{A.~V.~Khrabrov}
\email{skhrab@gmail.com}
\affiliation{Belle Mead, NJ 08502 USA}
\author{I.~D.~Kaganovich}%

\author{J.~Chen}
\altaffiliation[On leave from]{ Department of Engineering Physics, 
Tsinghua University,Beijing 100084, People’s Republic of China}
\affiliation{%
Princeton Plasma Physics Laboratory, Princeton NJ 08543 USA
}%

\author{H.~Guo}
\affiliation{Department of Engineering Physics, Tsinghua University, Beijing 100084, People’s Republic of China}

\date{\today}

\begin{abstract}
We study, by numerical and analytical means, the evolution of a uniform one-dimensional 
collisionless plasma initiated between plane absorbing walls. The ensuing flow is described by rarefaction waves 
that propagate symmetrically inward from the boundaries, interact, and eventually vanish after crossing through, 
leading up to the asymptotic phase.
Particle simulations indicate that the kinetic evolution qualitatively resembles
one well known in isentropic gas dynamics. Namely, a flattened density profile forms in the expanding central region where 
the propagating rarefaction waves interact, with a concomitant linear velocity profile. Asymptotically, the density 
falls off as $1/t$. Over the period when the rarefaction 
fronts still exist in the system, the density and the flux values at the boundary show only slight temporal variation. In gas dynamics,
these values would be exactly constant.
Evolution of the plasma potential, on the other hand, is strongly affected by the shape of the electron velocity distribution at $t=0$. 
If this distribution is Maxwellian, the potential drops off quite rapidly (on the underlying ion-acoustic time scale) to less than $T_e$ by the moment
when over $70\%$ of the initial plasma still remains in the system. This is due to electron kinetics being governed by conservation 
of adiabatic invariant in a slowly varying ambipolar potential well. Analytical model of the electron velocity distribution is presented 
to explain the simulations. The potential drop in the Debye sheath at the wall, as well as the ambipolar potential variation in the plasma, 
does not explicitly depend on the ion mass. The results may be of relevance to such applications as afterglow plasmas in pulsed discharges 
for material processing (ion implantation, isotope separation) and in gas-discharge switching devices. The formation of a flat density profile with 
$1/t$ decay rate may also be of use in applications requiring good plasma uniformity.
\end{abstract}

\maketitle

%


\section{\label{sec:Intro} Introduction}
\subsection{Prior work}
Collisionless decay of plasmas bounded by material surfaces, as well as a similar problem of plasmas with a sharp edge expanding 
into vacuum, has been a long standing subject of interest, e.g. in connection with laser-produced plasmas. 
Other established applications include expanding plasmas in astrophysics and planetary physics, including planetary wakes, 
as well as wakes behind orbiting satellites \cite{Samir83}.
The simplest case examined early on was that of a semi-infinite plasma \cite{Gur66}. Suppose the plasma is initially at $x>0$. 
Then in the quasineutral approximation, the flow at $x>0$ is in the form of a self-similar rarefaction wave propagating into the 
unperturbed state at the ion-acoustic speed $c_{s0}$. This solution is also valid if at $x=0$ there is an absorbing boundary \cite{allen70}. 
In that case, the rarefaction wave can be viewed as an expanding pre-sheath. In such flow, the Bohm condition at $x=0$ is automatically satisfied. 
For Boltzmann electrons and cold ions, the flow is identical to that of isothermal gas in a semi-infinite pipe when a piston on one end of it is 
being withdrawn at the speed of sound. 
An example of experimental observation of self-similar rarefaction flow is the work by Chung {\it et al.}\cite{Hershkowitz84}.
Certain kinds of non-equilibrium electron velocity distributions, e.g. bi-Maxwellian with sufficiently different 
temperatures, can result in a shock formation behind the rarefaction front \cite{Bezzerides78, wickens_allen_1979}, which itself is 
always a weak discontinuity. 

There is a also a large body of work on plasmas of finite size expanding into a vacuum 
\cite{Dorozhkina98,Mora03,Kovalev03,Mora09,Medvedev05,Murakami06}.
The aforementioned studies focus primarily on the edge structure of freely expanding plasma cloud where quasi-neutrality is 
violated and ions accelerate, and/or self-similar solutions valid for $t\ll L/C_{c0}$, where $L$ is the initial size of the plasma. 
The latter solutions do not carry information about the initial structure of the plasma cloud. 

Examples of recent work on evolution of bounded plasmas are in the areas of arc-plasma switches \cite{Sarrailh2008}, plasma-immersion implantation \cite{Chung2012} and ion extraction \cite{Chen2020}. All three works are studies of 
pulsed-plasma processes.
\subsection{Scope of the study}
In the present work, we concentrate on the expansion (decay) of a plasma confined between absorbing walls. The essential physics
of the process is similar to that of the inner flow in free expansion, especially at the initial stage when the electron distribution is not yet 
strongly modified by the shrinking ambipolar potential well. At the same time, the solution also involves a Debye sheath at the wall. 
In the quasineutral approximation we employ, the Debye sheath is represented by a potential jump the magnitude of which should be consistent 
with particle conservation. The electron kinetics of freely expanding finite plasma was previously studied by Mora and Grismayer\cite{Mora09} 
who pointed out the role of adiabatic modification of the electron velocity distribution. In their simulations, performed with adiabatic-Vlasov 
code, a flattening of the initial Maxwellian distribution was observed as the rarefaction waves moved in from the boundaries. Mora and Grismayer also
noted that the acoustic speed in the unperturbed state (where the ions are not yet moving), i.e. the speed of the rarefaction front, 
would change due to the adiabatic compression of trapped electrons. Those authors qualitatively characterized the flattened velocity distribution as a 
super-Maxwellian of the form $\exp\left(-(|v|/v_T)^n\right)$. Without account for adiabatic behavior, the plasma possesses a local equation of state 
and the quasineutral solution is given by gas dynamics. For Maxwellian electrons it was done, e.g., by Medvedev \cite{Medvedev05} who also compared the solution
with the result of a direct particle simualtion. 

Presently, using numerical (particle-in-cell) simulations as a starting point, we introduce a simple piecewise-linear parametrization of the evolving potential well.
It allows to give an explicit expression for the adiabatic invariant as a constant of motion, and therefore an expression for the electron velocity distribution function (EVDF). Such analytical expression is in good agreement with the EVDF found in simulations. We also bring attention to the fact that the kinetic evolution
of the decaying plasma still retains essential properties found in ordinary gas dynamics, and long-term asymptotic behavior corresponds to a gas
with the value of the adiabatic index $\gamma$ equal to $3$. Lastly, we direct particular attention at the time dependence of the plasma potential. Due to the 
flattening of the EVDF, the potential falls off to much lower values compared to that for a steady floating sheath, even as the fraction of the plasma remaining in the system is still large. Also, the potential scales with initial electron temperature, without a factor depending on the ion mass.

\subsection{Highlights}
It is already seen from general considerations that adiabatic modification of the EVDF is essential in the process at hand. At the initial stage, the ions (assumed cold) around the center plane will not start moving until the rarefaction front arrives, bringing a non-zero electric field. Therefore, the electron density near the center must also remain equal to the initial value $n_0$. On the other hand, in the collisionless regime high-energy electrons from the tail are being depleted (with resulting flux equal to that of the ions). The center density $n_0$ is maintained because the trapped electron population is compressed by the same rarefaction waves which cause acceleration of ions. The plasma potential sets at the value required to maintain quasineutrality. Once the wave fronts have passed through the center and an ion velocity profile has formed, adiabatic compression is succeeded by expansion. Combined with the continued tail depletion, this process returns the EVDF back to the initial shape, but cut off at a decreasingly smaller energy. Thus in the final, asymptotic phase, the EVDF can be considered as a flat-top. We recall and demonstrate that for a flat-top EVDF the plasma flow is identical to that of a gas with $\gamma=3$. The expansion process at the asymptotic stage is inertial with a flat density profile and a linear $x/t$ profile of ion velocity; the density falls off as $1/t$ and the potential as $1/t^2$. The case with initial flat-top EVDF is also simulated numerically, besides the Maxwellian case, to demonstrate that the known analytical solution is recovered. It should be noted that the $1/t$ density decay is universal in gas dynamics, except for isothermal case for which the leading-order time dependence is $1/(t \ln t)$.
\subsection{Structure of the article}
The article is organized as follows. Section~\ref{sec:model} presents the physical model and some aspects of its numerical implementation. 
In Section~\ref{sec:PIC}, numerical results are presented. To validate the numerical model, we begin with the case of a flat-top EVDF which is identical (apart from transient behavior) to rarefaction flow in $\gamma=3$ gas dynamics, with a known elementary solution. The Maxwellian case is considered next. Analytical model based on conservation of adiabatic invariant is introduced in Section~\ref{sec:analysis}. Quantitative analysis is given based on piecewise-linear approximation of the evolving ambipolar potential well. We also extend the comparison with rarefaction flow in isentropic gas dynamics (which is summarized for reference in the Appendix). The model, in particular, yields a quantitative expression for the EVDF that agrees with the numerical results. Conclusions and goals for future work are stated in Section~\ref{sec:conclusion}. 


\section{\label{sec:model} Physical and numerical model}
We consider a one-dimensional plasma composed of electrons and single-charged ions, initially at rest between plane absorbing boundaries at $x=0$ and $x=2L$. 
Due to symmetry, the boundaries can be treated as either floating or equipotential. Numerically, the resulting evolution is 
followed by means of a particle-in-cell simulation. Apart from the initial formation of Debye sheaths and/or transient oscillations, the characteristic time scale of the problem is $t_{0}=L/c_{s0}$, where $c_{s0}$ is the ion-acoustic speed in the uniform initial state. With time normalized by $t_0$, the evolution will not depend on the chosen value $M$ of the ion mass. In what follows, time will be given in normalized units, although different values of $M$ were used to verify the scaling. One can also use artificially low ion mass (say, a fraction of a.m.u.) to speed up the simulations. Most of the simulations presented here were performed with deuterium ions. 

Well-proven numerical code EDIPIC \cite{Sydorenko2006} was employed to perform particle-in-cell simulations of the plasma decay. It utilizes a semi-implicit algorithm for advancing the particles and solving the Poisson equation for the electrostatic potential.   
To follow long-term evolution of the system, sufficient spacial resolution and number of macroparticles need to be maintained at $t/t_{0}\gg 1$ as the plasma becomes strongly depleted while the Debye scale, as can be shown, decreases as $t^{-1/2}$. These considerations were taken into account. 

\section{\label{sec:PIC} Particle simulation results}
\subsection{Flat-top electon velocity distribution}
\begin{figure*}
\includegraphics[scale=0.5, angle=-90]{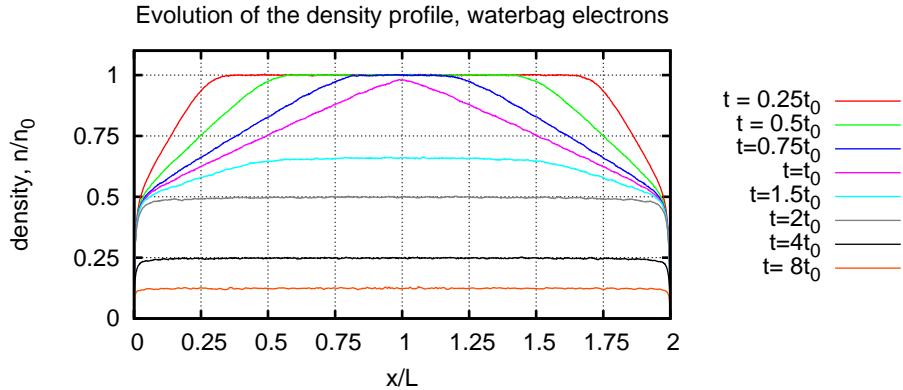}
\caption{\label{fig:DensityProfilesWaterbag} Density evolution for the case of waterbag electrons (EVDF is initially flat-top, making it such at all times). These results are in agreement with the $\gamma=3$ gas-dynamic solution (shown in Appendix) which applies in this case within the quasineutral approximation. Note the formation of a flat density profile with exact $1/t$ time dependence of the magnitude.}
\end{figure*}

We begin with presenting the results of a kinetic simulation for the case with a flat-top initial electron velocity distribution (also known as "waterbag" or "top hat"). In this case, the quasineutral kinetic model behaves exactly as a gas-dynamic model with the value of adiabatic index $\gamma=3$ equal to 3. Additionally, the $\gamma=3$ rarefaction flow problem has a straightforward analytical solution (given in the Appendix). Therefore, studying this case helps with overall understanding of the process and also allows to validate the simulation code. 
Figure~\ref{fig:DensityProfilesWaterbag} shows several successive density profiles for the plasma with flat-top electron distribution. The time scale $t_0$ equals $L/c_{s0}$, where $L$ is the half-size of the plasma. In this case, $c_{s0}=v_{\mathrm {max}}\sqrt{m/M}$, where $v_{\mathrm {max}}$ is the maximum (cut-off) velocity 
for the distribution at $t=0$.
 Rarefaction waves are seen to emerge from the boundaries and propagate with the velocity $c_{s0}$. The plasma density $n_w$ at the sheath edge (at the wall in the quasineutral approximation) remains unchanged over 
 the period $<0<t/t_{0}\leq t_{\mathrm cross}=2$, at the value equal to $n_w=n_0/2$. The time $t_{\mathrm cross}$ marks the instant when the rarefaction fronts have traveled all the way across to the opposite edge; in the case of flat-top EVDF their propagation speed does not change as the waves interact. The values of $t_{\mathrm cross}$ and $n_w$ depend on the initial EVDF but qualitatively, the evolution remains similar. We note that a flat density profile forms in the region between the two  rarefaction fronts after they pass through the center, with $n(t) = n_{0} (t_{0}/t)$ for $t\geq t_{0}$. These results agree with the analytical solution for $\gamma=3$ (valid in the quasineutral approximation) presented in the Appendix. 

Next in Fig.~\ref{fig:WaterBagFluxProfiles} we show successive profiles of the particle flux.
\begin{figure*}
\includegraphics[scale=0.45, angle=-90]{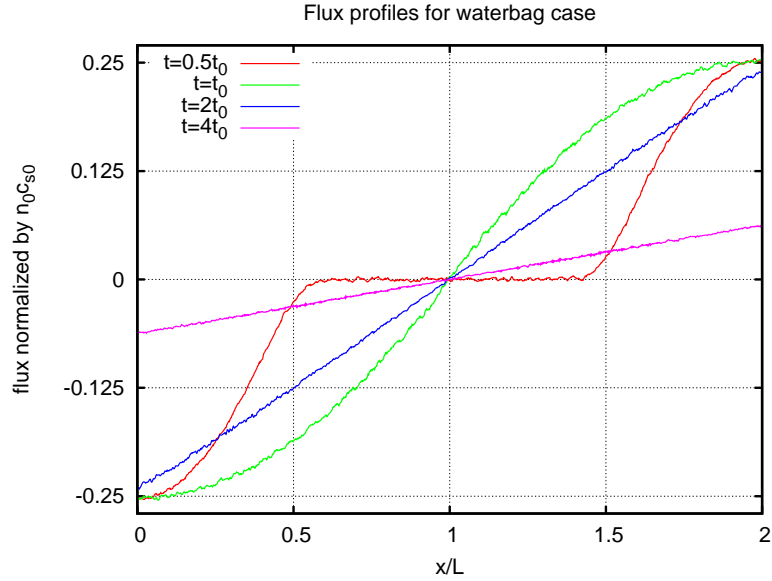}
\caption{\label{fig:WaterBagFluxProfiles} Successive snapshots of the flux profiles for the case with a flat-top initial distribution. Note the stationary sonic 
point at the boundary.}
\end{figure*}
As long as self-similar rarefaction-wave flows are still present in the vicinity of the walls for $t\leq t_{\mathrm cross}$, the wall flux remains constant, in this case equal to ($1/4)n_{0}c_{s0}$. At the boundary, the flux has a stationary point, consistent with maintaining constant density. The flow velocity at the boundary equals local acoustic speed, in this case $(1/2)c_{s0}$. 

For $t/t_{0}>2$, the flux falls off as $1/t^2$ because both the density and the velocity at the wall fall off as $1/t$. As the density profile becomes flat, the velocity profile becomes linear in $x$ of the form $x/t$. This is free inertial "red shift" decay. Such long-time asymptotic behavior, which does not depend on the initial EVDF, is an exact solution for the flat-top EVDF case.

\begin{figure*}
\includegraphics[scale=0.45, angle=-90]{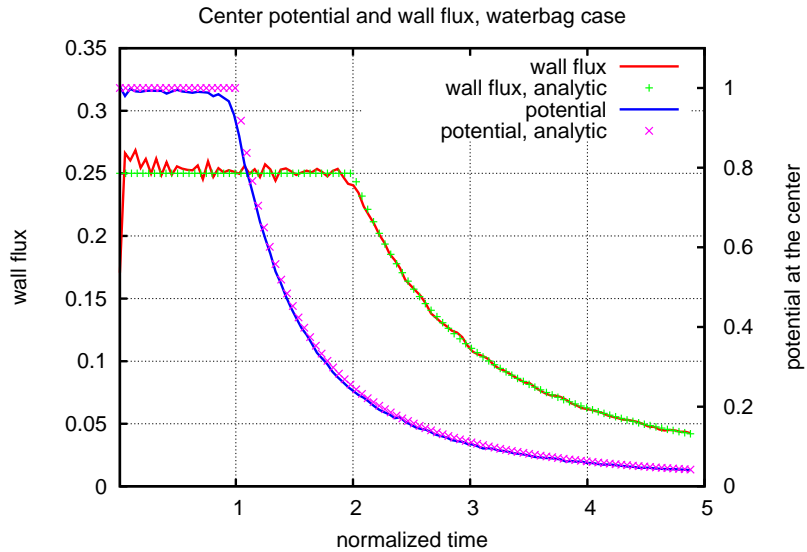}
\caption{\label{fig:FluxPotentialWaterbag} Time dependencies of the plasma potential and wall flux for the waterbag case. Good agreement with the gas-dynamic solution is observed.}
\end{figure*}

Time dependencies of the potential at the center $x=L$ and of the wall flux (at $x=0$ or $x=2L$) are plotted in Fig.~\ref{fig:FluxPotentialWaterbag}. 
Both analytical and numerical results are shown. A good agreement is observed between the two sets of data. 
Transient oscillations and are also seen in the numerical results, as a result of the initial formation of an ion-matrix and then the Debye sheath. 
While the flux evolution is qualitatively universal with respect to the initial EVDF (excluding unstable distributions with 
a negative $c_{s0}^2$, Eq.~(\ref{cs}), the constant value of the plasma potential over the period $t/t_{0}\leq 1$ 
is specific to the flat-top distribution case. Indeed, in this case the electron distribution depends on the potential
$\Phi$ locally (through the cut-off velocity) and therefore the collisionless plasma is characterized by a local equation of 
state with $\gamma=3$ ($n\propto \Phi^{1/2}$, $p \propto \Phi^{3/2}$). 

\begin{figure*}
\includegraphics[scale=0.45, angle=-90]{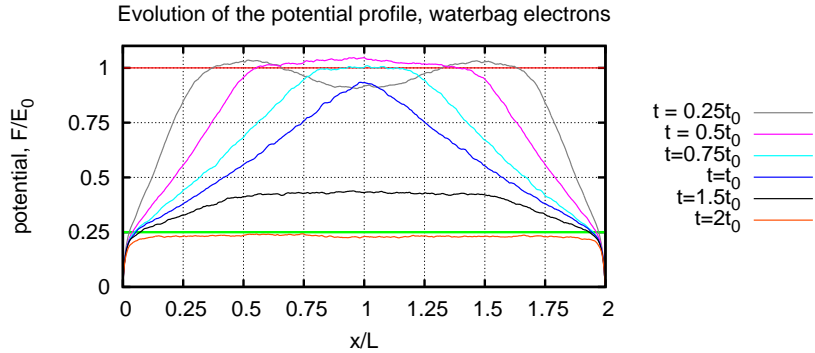}
\caption{\label{fig:WaterBagPotentialProfiles} Snapshots of the potential profiles for the flat-top initial EVDF. The normalization unit is the maximum electron energy at $t=0$. The horizontal lines at 1 and 0.25 the analytical for the respective moments of time in the quasineutral approximation.}
\end{figure*}

Lastly for the flat-top EVDF case, successive profiles of the potential in the plasma are shown in 
Fig.~\ref{fig:WaterBagPotentialProfiles}. 
Consistent with Fig.~\ref{fig:FluxPotentialWaterbag}, the potential in the unperturbed 
region remains constant in space (equal, in energy units, to the maximum energy of electrons at $t=0$). After the rarefaction fronts pass through the center, the potential profile in the wave interaction region becomes flat and falls off as $1/t^2$. There are also transient plasma oscillations seen to exist over a fraction of ion-acoustic time, forming standing waves. There is no Landau damping for the flat-top distribution. The amplitude is small compared to the ambipolar potential and the wavelength is large compared to the Debye scale. The effect of these oscillations on the ion motion can be neglected and thus quasineutrality still holds on the ion-acoustic time scale. It needs to be mentioned that formally, quasineutrality is violated at the rarefaction front, which in that approximation is a weak discontinuity where a $\delta$-function charge sheet is present. In particle simulations, as seen in the presented figures, the discontinuity is smoothed over many Debye scales with no large peaks in charge density. 

The case of a flat-top EVDF demonstrates sufficiently well the overall evolution of a decaying plasma bounded by walls. As stated, an analytical solution exists in the quasineutral approximation, making this case particularly useful for validating the simulation 
code. Next we proceed with the case of a Maxwellian distribution. The main interest, and the central point of the paper, is that the Maxwellian case shows non-trivial adiabatic evolution of the EVDF, unlike the flat-top case where the EVDF, as a function of the adiabatic invariant, is still a constant.

\subsection{Maxwellian electrons at $t=0$}
\begin{figure*}
\centering
\includegraphics[scale=0.45, angle=-90]{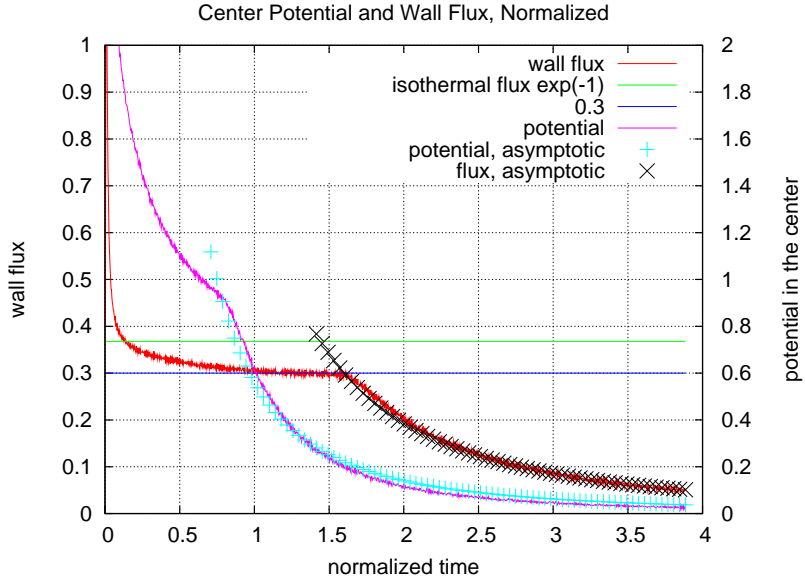}
\caption{\label{fig:FluxPotentialMaxwell}Time-dependent plasma potential (normalized by $T_e$, right y-axis) and wall flux (normalized by $n_{0}c_{s0}$, left y-axis) for the Maxwellian case. The behavior resembles that of a gas-dynamic solution, with slowly varying wall flux before entering inertial decay. The potential falls off to just below $T_e$ at $t=t_0$ when density $n_0$ is still maintained at the center and 70\% of the plasma remains in the system according to the corresponding density profile in Fig.~\ref{fig:DensityProfilesMaxwell}.}
\end{figure*}

\begin{figure*}
    \centering
    \includegraphics[scale=0.45, angle=-90]{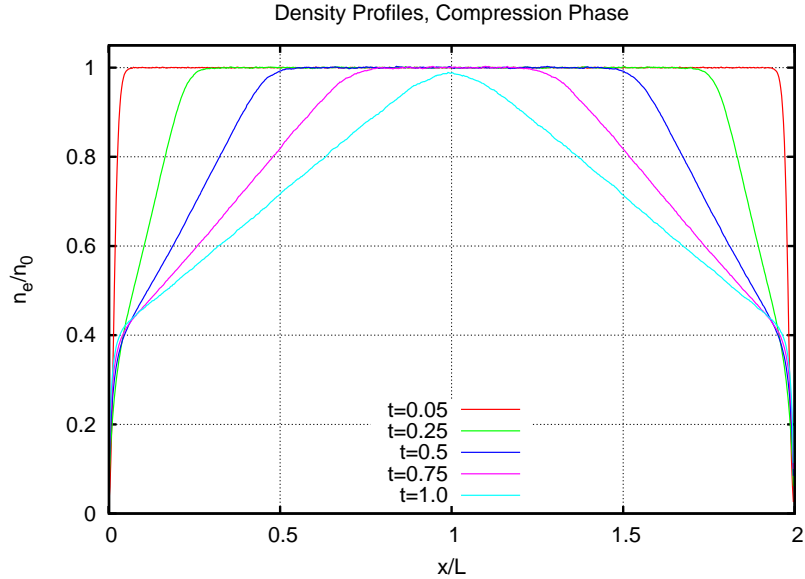}
    \caption{\label{fig:DensityProfilesMaxwell}Density profiles for a plasma with Maxwellian electrons for $t\leq t_0$, where $t_0$ is the observed time for the rarefaction front to reach the center. Note the slowly time-varying density value at the sheath edge of about $0.4n_0$.}
\end{figure*}

\begin{figure*}
    \centering
    \includegraphics[scale=0.45, angle=-90]{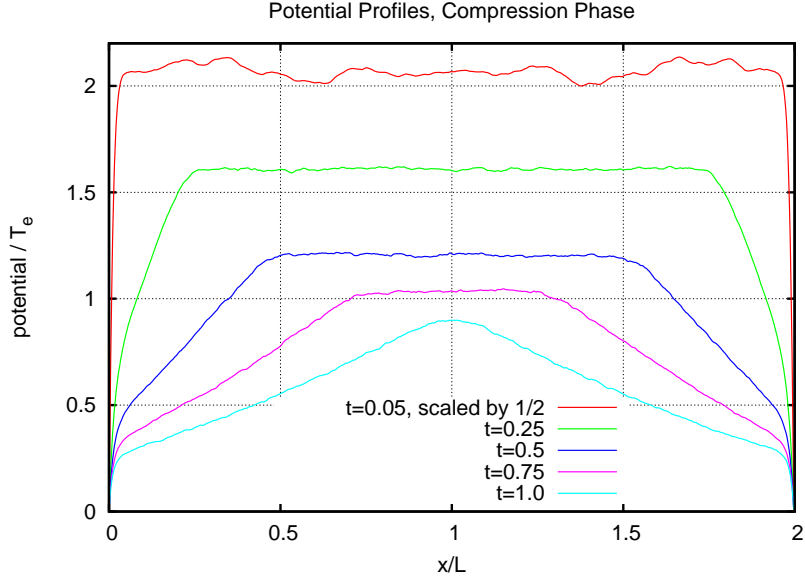}
    \caption{\label{fig:PotentialProfilesMaxwell}Potential profiles for the Maxwellian case. Only the initial profile depends on the ion mass. The Debye sheath potential is smaller than $T_e$ for each of the subsequent profiles shown, and the full potential falls to $0.9T_e$ at the instant of the rarefaction fronts meeting at the center.}
\end{figure*}

\begin{figure*}
    \centering
    \includegraphics[scale=0.45, angle=-90]{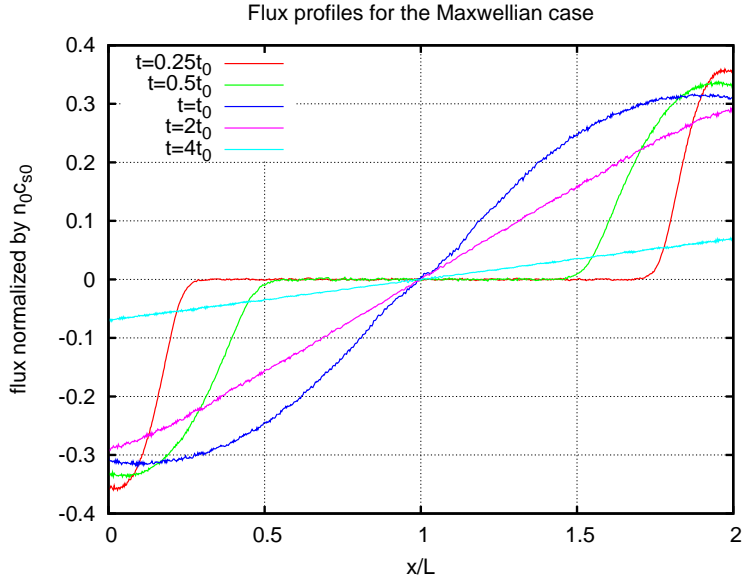}
    \caption{\label{fig:FluxProfilesMaxwell}Flux profiles for the Maxwellian case. Note that until the waves pass through, the flux has a stationary point near the wall, corresponding to the sonic point in a gas dynamic solution. This explains why the density at the wall changes slowly and the overall solution is structurally similar to propagation of self-similar rarefaction waves.}
\end{figure*}

Numerical results for the case with initial Maxwellian distribution of electrons are visualized in Figs.~\ref{fig:FluxPotentialMaxwell}--\ref{fig:FluxProfilesMaxwell}. Fig.~\ref{fig:FluxPotentialMaxwell} shows time dependencies of the plasma potential at the center and of the escaping flux, in the same format as Fig.~\ref{fig:FluxPotentialWaterbag} for the flat-top case.
The flux is very high over the short period of initial sheath formation, due to the presence of an energetic tail. The time of the formation of quasineutral rarefaction waves can be designated as the moment $t/t_0\approx 0.1$ at which the normalized flux equals $\exp(-1)$, the value found in isothermal gas dynamics (referenced in the Appendix). Between this moment and the time the rarefaction waves cease, the normalized flux varies within a narrow range between $\exp(-1)$ and $0.3$. This behavior is similar to isentropic gas dynamics in which the wall flux is initially constant. We note that the flow velocity is normalized by $c_{s0}=\sqrt{T_{e}/M}$ whereas the acoustic speed itself varies with time due to the adiabatic compression of trapped electrons. Indeed as seen in Fig.~\ref{fig:FluxPotentialMaxwell} from the variation of the plasma potential, the rarefaction waves reach the center $x/L=1$ at $t/t_0\approx 0.85$ and not at $t/t_0=1$ (variation of the potential will be addressed in more detail further on). The normalized time for the rarefaction fronts to cross over to the opposite wall is approximately twice that value at $1.7$. Thus the observed wave front traveling speed is approximately $1.2c_{s0}$. If the normalized flux at $t/t_0=1.7$ is re-scaled by this value, it will be equal to $0.25$ which actually corresponds to the case $\gamma=3$ considered in the previous section. However, an effective value of the adiabatic index is difficult to assign because the shape of the EVDF, which determines it, varies both in time in space. For example, the snapshots of density profiles, plotted in Fig.~\ref{fig:DensityProfilesMaxwell}, show that the plasma density at the sheath edge stays at the value of approximately $0.4n_0$. We note that in a numerical study \cite{Sarrailh2008} of the post-conduction phase in a 
plasma-arc switch such value of density at the sheath edge was indeed observed on the anode side. A flat density profile with $n=0.4n_0$ was
seen to form and then gradually decay.
For a gas-dynamic rarefaction wave, the corresponding value of the adiabatic index $\gamma$ would be approximately $1.4$. Note that in figures \ref{fig:DensityProfilesMaxwell} to \ref{fig:FluxProfilesMaxwell}, unlike previously, the time argument is normalized by the actual time for the wave front to travel half-length $L$ of the bounded plasma. For clarity, the snapshots cover only this phase of the plasma decay and not the subsequent interaction of the rarefaction waves. This interval is also sufficient to discuss the adiabatic evolution of the EVDF in the following section, since at later times, as the rarefaction fronts move apart, this evolution is simply reversed and the EVDF becomes progressively closer to a cut-off Maxwellian.

We now return to the time variation of the plasma potential, plotted in Fig.~\ref{fig:FluxPotentialMaxwell}. The important fact is
that at the moment when the
rarefaction fronts meet at the center, the plasma potential already falls to the value of $0.9T_e$ an this value is 
sufficient to maintain the electron density 
at the center still equal to $n_0$. Besides being slightly smaller than $T_e$, this value also does not depend 
on the ion mass $M$. At this moment, approximately $70\%$ of the plasma is still left in the system, as seen from Fig.~\ref{fig:DensityProfilesMaxwell}. 

Successive snapshots of the potential profiles in the Maxwellian plasma with rarefaction waves are shown in Fig.~\ref{fig:PotentialProfilesMaxwell}.
It is seen that the potential profiles between the sheath edge and the rarefaction front and are approximately linear, 
Such would actually have been the case for an isothermal gas (i.e. plasma with Boltzmann electrons), even though the observed potential 
variation is not consistent with isothermal rarefaction wave where it equals to $T_e$). 
This observation will be utilized in section \ref{sec:analysis} to propose a piecewise-linear approximation of the potential well.

For the plasma flux profiles, the snapshots are shown in Fig.~\ref{fig:FluxProfilesMaxwell}. In similarity with the previously discussed $\gamma=3$ case and isentropic gas dynamics in general, over the period when the rarefaction waves are present, the wall flux shows slow variation in time and there is a stationary point located at the wall or in close vicinity. At larger times, $1/t^2$ asymptotic dependence is seen for the wall flux: the value at $t=4$ is $1/4$ of the value at $t=2$.
Finally, the asymptotic behavior of both the wall flux and plasma potential is the same as seen in the case of a flat-top EVDF, namely both fall off as $1/t^2$. No data fitting was applied to produce the asymptotic curves; each is based on a single data point from the simulation. We can now proceed with analyzing the simulation results.

\section{\label{sec:analysis} Analysis}
To analyze the particle simulation results, we utilize an easily tractable model with piece-wise linear profile of the potential. The adiabatic invariant can be easily calculated in this case. A fully self-consistent solution is not presented, although the required set of equations is specified. The simplified model of the potential well is sufficient to gain good insight into the process and, in particular, obtain a quantitative expression for the EVDF that shows good agreement with the simulation data. Also, the resulting analytical expressions for the local plasma density and for the total number of particles (per unit area of the wall) allow, in principle, to impose constraints on the parameters specifying the potential profile at a given state of the system. 

We introduce the following notation: $L$ is half-width of the plasma slab, $X=X(t)$ is the position of the rarefaction front relative to the boundary (before or after passing through the center), $\Phi=\Phi(x,t)$ is the electrostatic potential and $U=-e\Phi$ is the electron potential energy, with $U=0$ at the center $x=L$ of the decaying plasma. The time-depending Hamiltonian is $H=U(x,t)+mv^2/2$. The symmetric potential $U(x,t)=U(2L-x,t)$ is assumed to be flat in the region between the rarefaction fronts (before or after they cross the middle plane), vary linearly in the quasineutral rarefaction wave, and also have sheath jumps at the boundaries $x=0$ and $x=2L$, with negligible width. Specifically, to the left side $0<x<L$ of the symmetry plane $x=L$,  
\begin{equation}
  U(x,t) = \left\{
    \begin{array}{rl}
                        0, &   X(t) < x \leq L,\\
      U_w(t)\left[1-x/X(t)\right], &  0<x\leq X(t),\\
                     U_{max}(t),&  x=0.
    \end{array} \right. \label{eq:potential}
\end{equation}
It is seen that $U_{max}$ is the depth of the potential well and $U_w$ is the ambipolar potential variation in the plasma. The sheath potential is $U_{max}-U_w$. Electrons trapped in the potential (\ref{eq:potential}) with energies above $U_w$ but below $U_{max}$ bounce off the negligibly thin potential barrier presented by the wall sheath. 

\begin{figure*}
    \centering
    \includegraphics[scale=0.45, angle=-90]{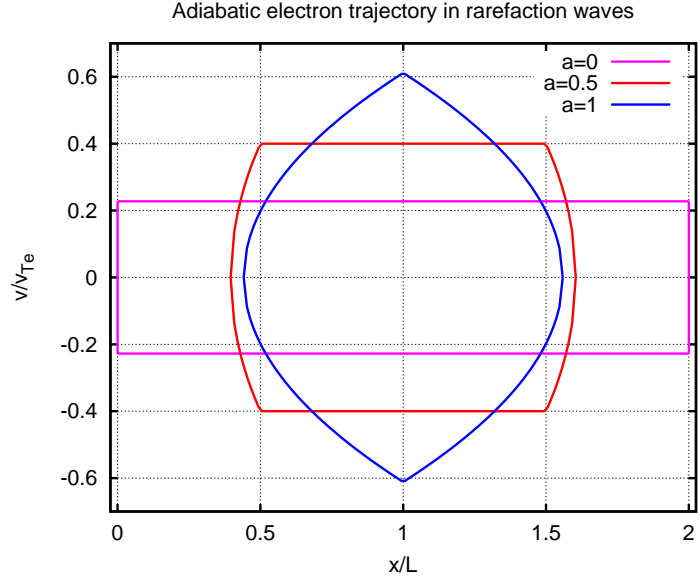}
    \caption{\label{fig:cartoon}Illustration of the adiabatic evolution in one-dimensional phase space. In the potential (\ref{eq:potential}), the orbits are composed of straight and parabolic segments and there are also reflections at the step representing the sheath. The enclosed area
    is preserved under the mapping. The parameter $a$ is the relative distance $X(t)/L$ traveled by the rarefaction fronts towards the center.}
\end{figure*}

\begin{figure*}
    \centering
    \includegraphics[scale=0.6, angle=-90]{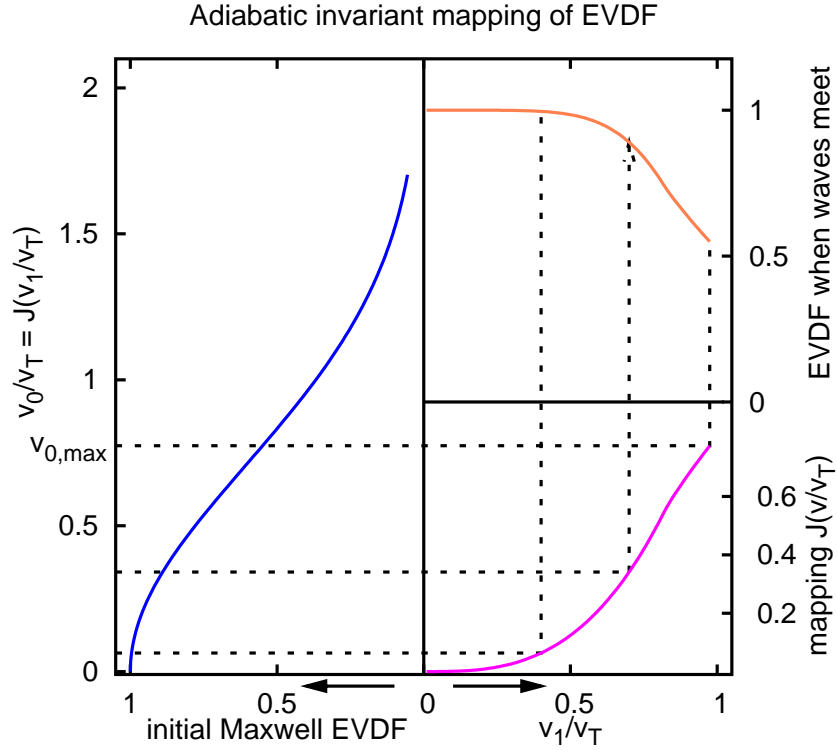}
    \caption{\label{fig:EVDFmap}Mapping of the EVDF defined by the conservation of adiabatic invariant $J$. In the initial state with a Maxwellian distribution, the normalized invariant $J$ equals $v/v_T$. The function $J(v/v_{T})$ for $a=1$, with rarefaction fronts meeting at the center, is shown in the bottom right quadrant. The resulting transformation of the Maxwellian EVDF yields the distribution shown on the top. Expressed as a function of total energy, this distrubution applies in the entire domain.}
\end{figure*}

\begin{figure*}
    \centering
    \includegraphics[scale=0.45, angle=-90]{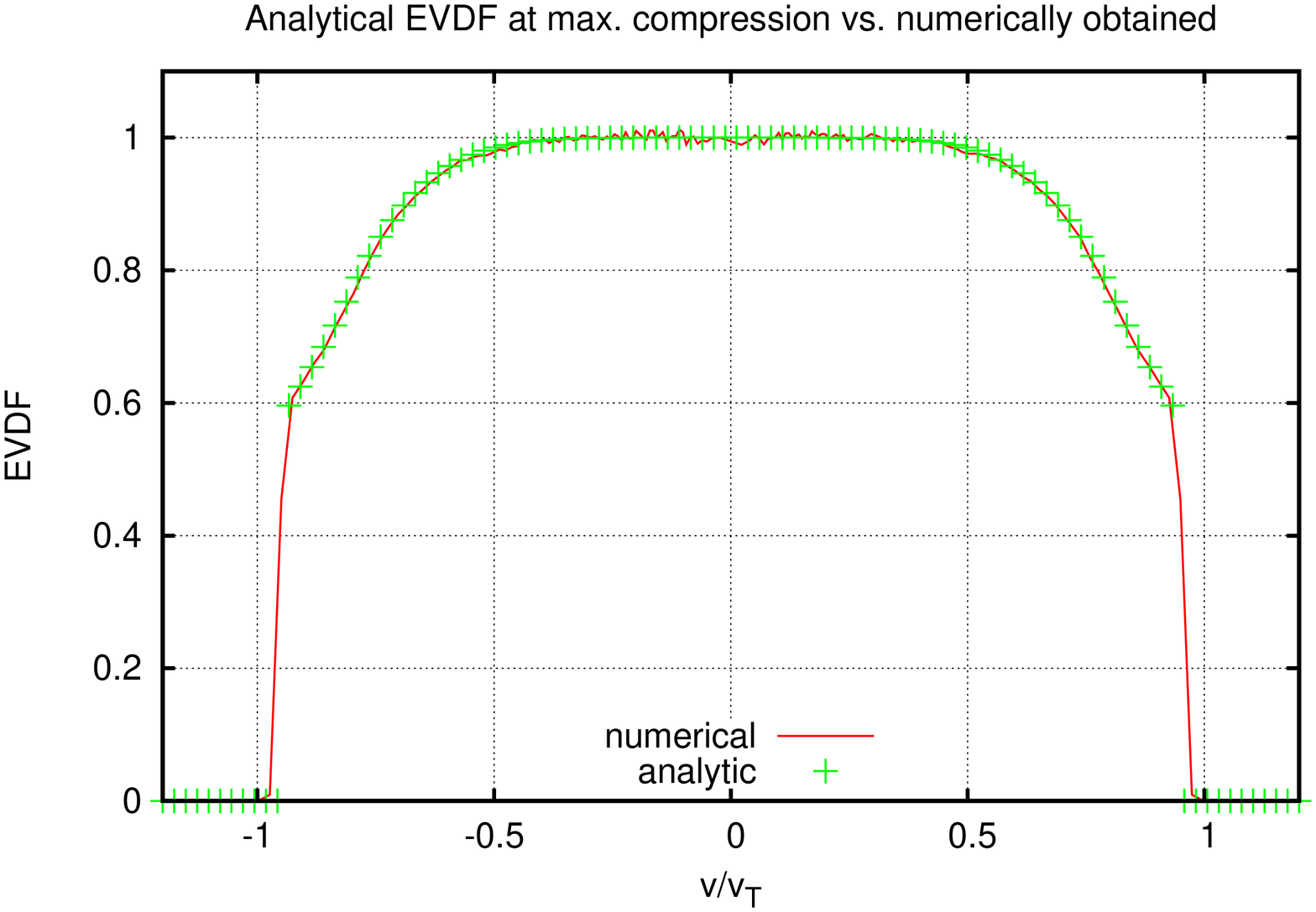}
    \caption{\label{fig:EVDFcompare} Comparison between analytical and numerical EVDFs. No fitting was performed; the parameters of the potential well were calculated based on the observed maximum value and the density at the sheath edge. At the moment when the wave fronts meet at the center, such calculation gives $u_{max}=0.95$ and $u_w=0.67$. These are, in units of $T_e$, the potential maximum and its variation along the rarefaction wave. Thus the Debye sheath potential is $0.28T_e$.}
\end{figure*}

Since the shape and the magnitude of the ambipolar potential vary on the ion-acoustic time scale which far exceeds the bouncing time of electrons, the adiabatic invariant 
\begin{equation}
I=I(H)=\oint vdx=\oint\sqrt{\frac{2}{m}\left(H-U(x,t)\right)},
\end{equation}
where the integral is taken over a bouncing period, is conserved, and so is the electron velocity distribution function 
$f_{v}=f_{v}(I(H))=f_{v}\left(I(\frac{mv^2}{2}+U)\right)$. The subscript "$v$" indicates that the normalization is defined by integration
over velocity. Conservation of adiabatic invariant for trapped electrons was originally considered by Gurevich \cite{Gur68}.
We recall that $I(H)$ equals the phase area enclosed by the given orbit.
In particular, the number of confined particles (per unit area of the wall) with energy values up to $H$ is given by 
\begin{equation}
    N(H)=\int_{0}^{I(H)}f_{v}(I^{\prime})dI^{\prime}.
    \label{normalize1}
\end{equation}
Fig.~\ref{fig:cartoon} illustrates adiabatic evolution of an electron orbit. The trajectories correspond 
to three different positions of the rarefaction fronts, showing energy gain as the fronts travel inward.   
The velocity distribution $f_v$ expresses explicitly as a function of $I$ if at the initial moment the potential well is 
rectangular (will variation in the negligibly thin wall sheath only): 
\begin{equation}
f_{v}(I)=f_{v,0}\left(\frac{I}{4L}\right).
\end{equation}
To proceed further, let us introduce the scaling parameters and non-dimensional variables as follows:
$u=\frac{U}{T}$, $h=\frac{H}{T}$, $u_{max} = \frac{U_{max}}{T}$, $u_{w} = \frac{U_{w}}{T}$, $T = \frac{mv_{T}^2}{2}$, $\tilde{n}(x,t)=\frac{n}{n_0}$, $a=a(t)=\frac{X(t)}{L}$. 
Here, $v_T$ is the characteristic velocity for the initial distribution and $T$ the corresponding energy. For a Maxwellian distribution, $v_T$ can be taken as the thermal velocity, with $T=T_e$ as a result, but the treatment is not restricted to the Maxwellian case. The adiabatic invariant is normalized as $J =\frac{I}{4v_{T}L}$ and the velocity distribution is given as
\begin{equation} \label{evdf}
    f_{v}(I)=\frac{n_0}{v_T}g\left(J(h,a,u_w)\right),
\end{equation}
where 
\begin{equation}
2\int_{0}^{\infty}g(J)dJ=1.
\label{normalize2}
\end{equation}
Note that the two normalizations (\ref{normalize1}) and (\ref{normalize2}) are consistent.
The dependence upon the potential maximum $U_{max}$ is omitted in Eq.~(\ref{evdf}); it is in the form of a step function. 
For the potential given by Eq.~(\ref{eq:potential}), the normalized invariant $J$ expresses as follows:
\begin{equation} \label{J_of_h}
    J(h,a,u_w)=\sqrt{u_w}\tilde{I}\left(\frac{h}{u_w}, a\right),
\end{equation}
where $h<u_{max}$ and 
\begin{equation} \label{I_of_eps}
\tilde{I}(\epsilon,a)=\epsilon^{1/2}(1-a)+\frac{2}{3}a\left(\epsilon^{3/2}-\Theta(\epsilon-1)(\epsilon-1)^{3/2}\right).
\end{equation}
In the above expression, $\epsilon$ is the ratio $H/U_w$ of the electron energy $H$ to the ambipolar potential variation $U_w$ and $\Theta$ is the Heaviside step function.
The velocity distribution, as a function of the total energy $h$ in the potential well at a given time (on the "slow" ion-acoustic scale of quasineutral evolution), expresses via Eqs.~(\ref{evdf}), (\ref{J_of_h}), and (\ref{I_of_eps}).
The mapping giving EVDF for a given shape of the potential well is illustrated in Fig.~\ref{fig:EVDFmap}. 
Fig.~\ref{fig:EVDFcompare} shows a comparison of the above distribution with the one computed from simulation data.  
The presented case is for $a=1$, that is the moment for the rarefaction fronts to reach the center. The known values of the density at the center and at the wall (the latter from simulations) impose, through Eq.~\ref{density} below, two conditions on the ambipolar potential $u_w$ and on the total potential $u_max$ both of which can thus be determined. Those values, which also agree with the simulations, were used in calculating the predicted distribution according to Eq.~\ref{evdf}. Also, the fraction of the plasma remaining in the system is consistent with that predicted by Eq. ~(\ref{normalize1}). The latter specifies to $N=\mathrm{erf}\left(J(u_{max},a,u_w)\right)$ in accordance with Eqs.~(\ref{normalize1}) and (\ref{J_of_h}).

We introduce the following non-dimensional expressions for the electron density, the local acoustic speed $\tilde{c_s}$, and the pressure $\tilde{p}$. The acoustic speed is normalized by $\frac{1}{2}\frac{M}{m}v_T^2$, where $M$ is the mass of the ion species, and the pressure is normalized by $n_{0}T=\frac{1}{2}mv_T^2$.
\begin{equation}  \label{density}
    \tilde{n}(u,a,u_w,u_{max})=\int_u^{u_{max}}dh (h-u)^{-1/2}g(J(h,a,u_w)),
\end{equation}
\begin{equation}    \label{cs}
\begin{split}
\tilde{c_s^2}(u,a,u_w,u_{max}) & = \\
   & \frac{\tilde{n}(u,a,u_w)}{\frac{g(u_{max},a,u_w)}{(u_{max}-u)^{1/2}}-\int_u^{u_{max}}\frac{dh}{(h-u)^{1/2}}\frac{dg}{dJ}\frac{\partial J}{\partial h}},
    \end{split}
 \end{equation}
 \begin{equation}\label{pressure}
    \tilde{p}(u, a, u_w, u_{max})= 2\int_u^{u_{max}}dh(h-u)^{1/2}g(h, a, u_w).
\end{equation}
In Eqs.~(\ref{density})--(\ref{pressure}) the $x$-dependence is through the potential $u=u(x,a,u_w)$ as defined by Eq.~(\ref{eq:potential}). Eq.~(\ref{cs}) is the known general expression for ion-acoustic speed in the long-wave limit. The energy derivative of the distribution is taken by applying the chain rule. Note that $\partial J/\partial h$ is the normalized bouncing period, obtained by differentiating Eq.~(\ref{J_of_h}). The boundary contribution ($\delta$-function term) is present in Eq.~(\ref{cs}) due to the jump of $g$ to zero at $u=\pm u_{max}$. Such terms, arising in wave dispersion equations for cut-off distributions, are sometimes not accounted for in published studies.

Differentiating Eq.~(\ref{pressure}) with respect to $x$ shows that for electrons the pressure is balanced by the electric field.
This condition was expected although not used explicitly.

We note briefly that Eq.~(\ref{density}) provides a non-local quasi-neutral closure to the standard fluid equations for cold ions:
\begin{equation}
\begin{array}{rl}
    \frac{\partial n}{\partial t} + \frac{\partial}{\partial x}\left(nV\right)= 0, \\
    \frac{\partial V}{\partial t} + V\frac{\partial V}{\partial x} = -\frac{e}{M}\frac{\partial\Phi}{\partial x}.
    \end{array}
\end{equation}
An applicable numerical scheme would use some type of iterative solution for the nonlinear integral equation (\ref{density}) to find
the potential profile (including the jump at the wall) at each step and then advancing the ions in the electric field. Ions in such scheme could also be 
treated as particles if their thermal spread needs to be accounted for. This topic will not be pursued at the present time.
If the above equations are applied to the case of a flat-top initial velocity distribution (the scaling parameter $v_T$ 
can be chosen as the maximum velocity), in which case $g(J)$ is also flat, it it seen that the ion fluid behaves as a gas with 
$\gamma=3$ and therefore the quasineutral evolution is governed by the respective set of equations 
of gas dynamics, with electrostatic potential, as a result, proportional to density squared. The gas dynamic solution is 
addressed in the Appendix. The main properties of the gas-dynamic solution are as follows. Initially, it is in the form of self-similar rarefaction waves propagating symmetrically inwards from the boundaries. The ions at any given position begin to 
move after the rarefaction front passes through. The density and the flow velocity (equal to the local acoustic speed) at the wall remain constant; for $\gamma=3$ they both equal $1/2$ of the respective value in the unperturbed state. After the rarefaction fronts pass through the center (which can also be viewed as reflection of the rarefaction waves), a flat density profile and a linear velocity profile form in the expanding central region. For $\gamma=3$ this behavior is exact, but the flow is qualitatively similar for other values of the adiabatic index. After the wave fronts have traveled all the way across, the density profile becomes flat (at $1/2$ the initial value for the flat-top EVDF, or approximately $0.4n_0$ in the Maxwellian case) and then decays inertially as $1/t$. The flux to the boundary remains constant up to the moment when the wave fronts cross the plasma slab and falls off as $1/t^2$ afterwards. So does the plasma potential. Such asymptotic behavior is universal because the electron distribution becomes cut off at velocities much lower than thermal, that is, it becomes a flat-top distribution in the long run.

\section{\label{sec:conclusion} Conclusions and future work}
We have investigated, numerically and analytically, collisionless rarefaction flow of a plasma bounded by planar walls. Qualitatively, the evolution of the system in certain ways resembles that in gas dynamics. Specifically, rarefaction waves 
are launched from the boundaries and interact in the center region. The density and the flux at the wall show little variation in time until the rarefaction waves pass across and cease to exist. A flat density profile forms in the center and extends to the boundaries as the rarefaction fronts pass travel through. At later times, the density decays as $1/t$ and the flux falls off as $1/t^2$. The kinetic description of the process is based on the conservation of adiabatic invariant for electrons that are trapped in the system at any given moment. An interesting consequence of such evolution is the resulting low value of the potential in the plasma, for example $0.9T_e$ at the moment when the rarefaction fronts reach the center of the plasma slab. At that instant, the density at the center still equals the initial value and 70\% of the plasma still remains in the system. The numerical model and analysis of the results can be applied to an arbitrary initial distribution of electrons (as long as $c_s^2>0$).
The formation of a flat density profile and low plasma potential may be of interest in material processing applications where uniformity is important and low ion energies are often desired. One example of a process where uniformity is essential is Raman amplification of short laser pulses where the required tolerance is within few percent \cite{Ping2003}. 

At present, we have not developed a fully self-consistent solution for the quasineutral adiabatic model of the rarefaction flow, but presented an analysis which predicts the EVDF in good agreement with simulations. Developing a fully self-consistent solution will be a subject of future work. The task at hand would be to formulate a scheme for solving the corresponding set of integrodifferential equations (hydrodynamic equations for the ions with quasineutrality closure based on the adiabatic EVDF).

\section{Data availability statement}
The data that support the findings of this study are available from the corresponding author upon reasonable request.

\appendix
\section{Rarefaction flow in gas dynamics}
The purpose of this section is to aid in interpeting the results of our particle simulations. Basic understanding of the decay of a bounded plasma can be gained from considering a gas-dynamic (fluid) approximation of the problem with appropriate boundary conditions. Specifically, the outflow velocity should equal the local acoustic speed at the sheath edge. On the quasineutral time scale, assuming the Debye sheath to form instantaneously at $t=0$, the initial phase of the process is a self-similar rarefacton wave. Such solution in the case of plasma was originally obtained for a semi-infinite configuration with Boltzmann electrons (corresponding to isothermal gas with electrostatic potential playing the role of enthalpy) \cite{Gur66}. In a finite plasma bounded by planar walls, the ion-flow regions also spread inward initially as rarefaction waves. The two rarefacton fronts meet at the symmetry plane and pass through each other, forming an expanding region where the rarefaction waves interact. The solution in the interaction region is relatively simple to obtain for specific rational values of the adiabatic index $\gamma$ \cite{Landau87} and is also known in general form \cite{riemann1860,Jenssen17}, less suitable for numerical calculation. An analytical solution for $\gamma=2$ was given, for example, by Startsev \cite{Startsev04} who also noted the $1/t$ asymptotic time dependence of the density field.
In this section, a full solution will be given, for reference, only for a gas with $\gamma=3$, in which case it is elementary and at the same exact for a flat-top initial electron distribution as stated in the main text.
Further, the long-time asymptotic behavior ("red-shift decay") such solution displays is correct for the initial electron velocity distribution of sufficiently general shape (e.g. monotonic with a smooth maximum at zero velocity).
Since a specific value of the adiabatic index cannot be assigned to a plasma with adiabatically evolving electron distribution, we also plot gas-dynamic solutions for $\gamma=1$ and $\gamma=2$. In the kinetic model with Maxwellian EVDF, the effective adiabatic index calculated as $c_s^{2}(\rho/p)$ is always in the range between 1 and 3 (with spatial and temporal variation); therefore showing the gas-dynamic solutions should help with qualitative understanding of the process.

\begin{figure}
    \centering
    \includegraphics[scale=0.7, angle=-90]{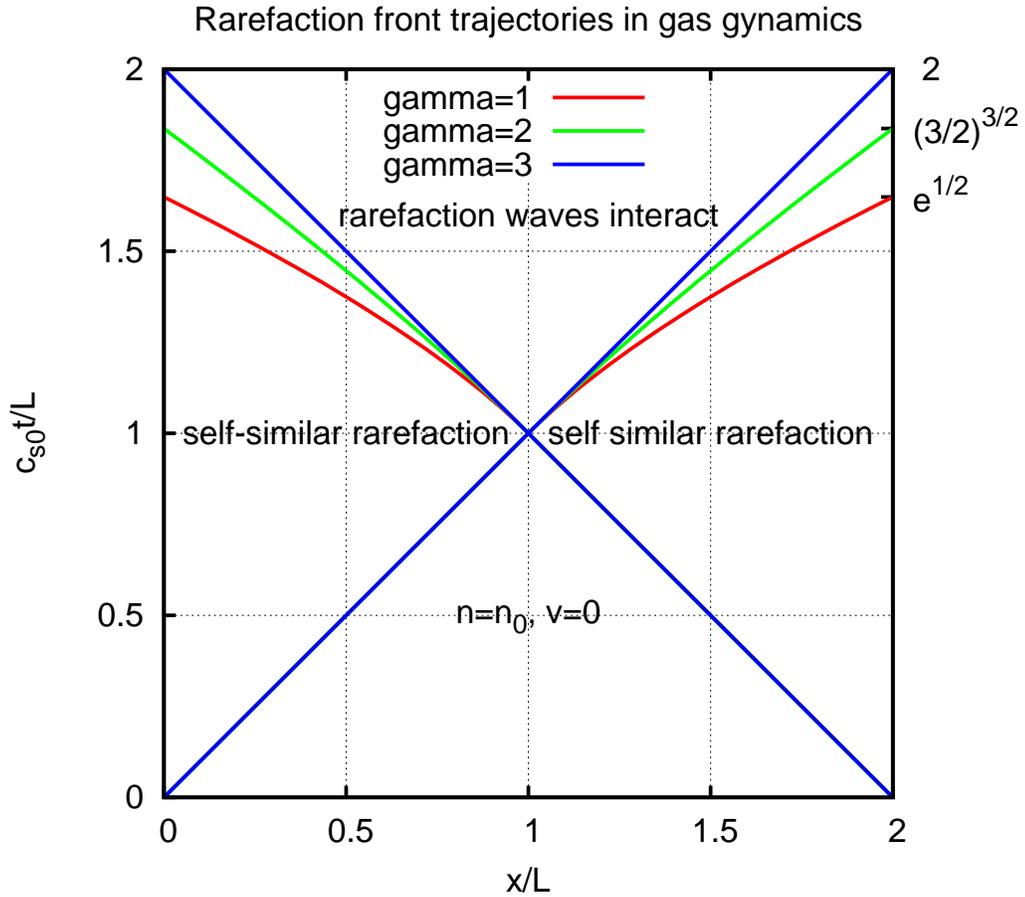}
    \caption{Rarefaction front trajectories for $\gamma = 1, 2, 3.$ For plasma with adiabatic electrons, the acoustic speed in the center is changing even when ions at $x=1$ are still at rest.}
    \label{FrontTraj}
\end{figure}

In what follows, the gas with adiabatic index $\gamma$ is assumed to be initially at rest within a slab region $0<x<2$ with initial density and acoustic speed both equal to unity. The rarefaction fronts propagate into the unperturbed gas with velocity $1$. For $t\leq 1$, the solution in the left half $0<x<1$ for velocity $u(x,t)$ and density $n(x,t)$ has a self-similar form, with $\xi=x/t$:
\begin{equation}
    u(x,t,\gamma) = \frac{2}{\gamma+1}\left(\xi-1\right),
\end{equation}
\begin{equation}
    n(x,t,\gamma) = \left[\frac{2+(\gamma-1)\xi}{\gamma+1}\right]^{\frac{2}{\gamma-1}}.
\end{equation}
The solution for $\gamma=1$ is found by taking a respective limit and yields
\begin{equation}
    n(x,t,1)=\exp(\xi-1),
\end{equation}
\begin{equation}
   u(x,t,1)=\xi-1. 
\end{equation}
The ambipolar potential $\Phi(x,t)$, in units of $p_{0}/n_0$, is found from the pressure balance for electrons and it varies as $n^{\gamma-1}$, apart from an additive constant. The isothermal case (i.e. Boltzmann electrons) yields a familiar result $\Phi\propto\ln{n}$. Note that in the gas-dynamic model, the potential in the unperturbed region for $1<t$ remains constant because it is a local function of $n$.
Another useful property of the rarefaction-wave flow is the relation between the velocity and local acoustic speed:
\begin{equation}
    c_s = 1 - \frac{\gamma-1}{2}|u|.
\end{equation}
The gas at $t<x<2-t$ remains unperturbed. At $t>1$, the rarefaction fronts will travel past the center plane. The front velocity no longer equals the equilibrium value of $1$ and the trajectory can be found by the method of characteristics \cite{Landau87} (in terms of which, the trajectory is a line separating two types of flow). For $1\geq\gamma\le 3$, the wave front accelerates at $t>1$; for $gamma=3$ it continues to travel at the constant speed of 1. The moment for the wave front to reach the opposite boundary depends on $\gamma$; trajectories are plotted in Fig.~\ref{FrontTraj}.
Until the wave fronts have traveled all the way across, the flow next to the boundary is still given by the self-similar solution at $\xi=0$ for a given $\gamma$. The velocity $u$ (equal to the local acoustic speed), the density $n$, and the wall flux $\Gamma=nu$ maintain constant values.  Also note that at the boundary $\frac{\partial\Gamma}{\partial x}=0$. The solution in the interaction region for $t>1$ is especially simple in the case of $\gamma=3$: the density between the spreading wave fronts equals $1/t$ and the velocity profile is linear, matching up to the self-similar solutions outside. At $t>2$, the density profile becomes constant (flat) and continues to fall off as $1/t$. The potential, accordingly, varies as $1/t^2$. As an asymptotic solution, such behavior is universal for collisionless systems evolving inertially. The solutions for $\gamma=1$ and $\gamma=2$ are qualitatively similar to the $\gamma=3$ case up until the moment the wave fronts cross the plasma. The $1/t$ asymptotic decay of the density is also reproduced, although not the $1/t^2$ asymptotic decrease of the sheath potential. In what follows, we illustrate the gas-dynamic solutions for $\gamma=1, 2, 3$. The inner solutions for $\gamma = 1, 2$ in the interaction region are obtained by a suitable approximation based on the conservation of mass. The density profile is parabolic and the velocity profile is linear. This approximation works quite well for a bounded plasma where the scale length does not vary with time. The plotted curves are indistinguishable from those representing exact solutions.
\begin{figure}
    \centering
    \includegraphics[scale=0.7, angle=-90]{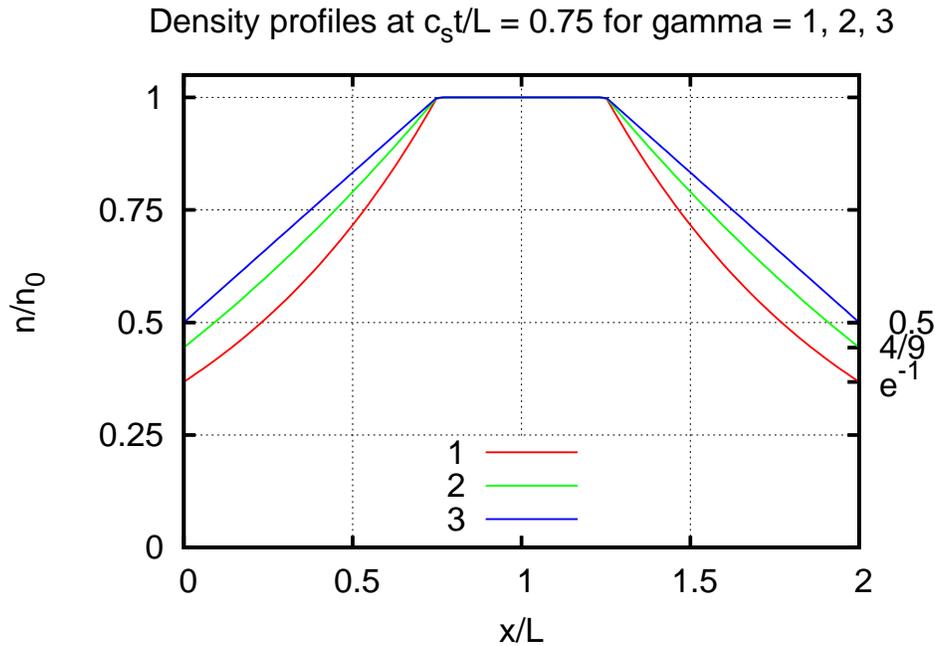}
    \caption{Density profiles at $t=0.75L/c_{s0}$ for three values of the adiabatic index $\gamma$. Note the boundary values.}
    \label{DensAt075j}
\end{figure}
\begin{figure}
    \centering
    \includegraphics[scale=0.6, angle=-90]{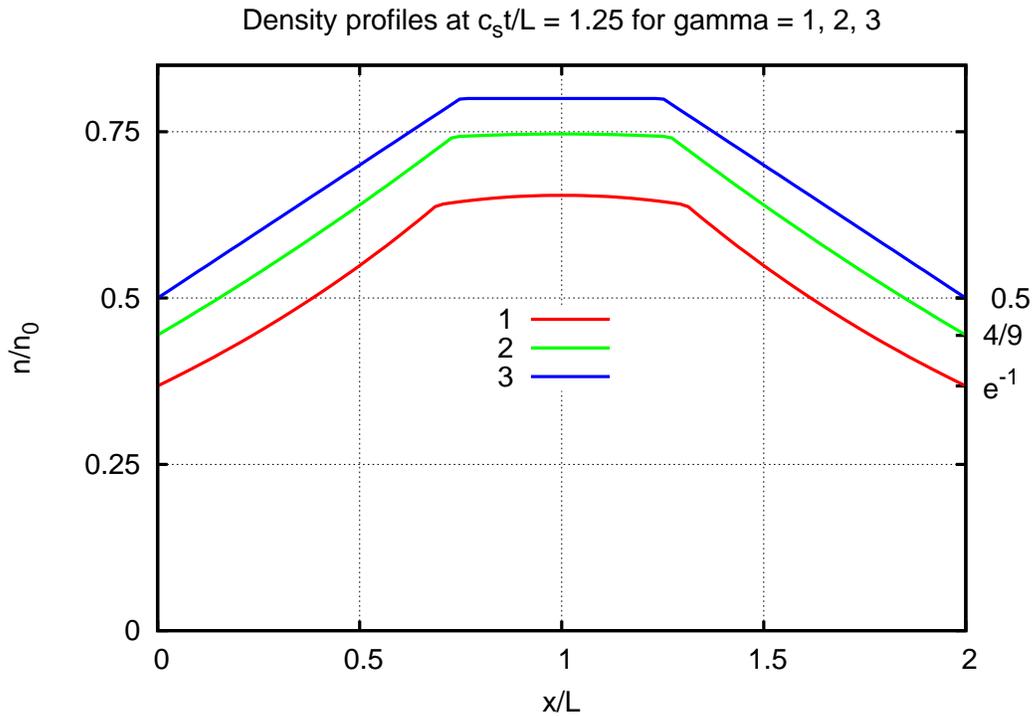}
    \caption{Density profiles at $t=1.25L/c_{s0}$ for three values of $\gamma$. Note the decay at the rate of $1/t$ or faster. Asymptotically, the rate is $1/t$ for $\gamma>1$ and $1/(t\ln t)$ for $\gamma=1$. The curvature of the profile is small in the interaction region; this property is retained in the kinetic solution.}
    \label{DensAt125}
\end{figure}    
\begin{figure}
    \centering
    \includegraphics[scale=0.7, angle=-90]{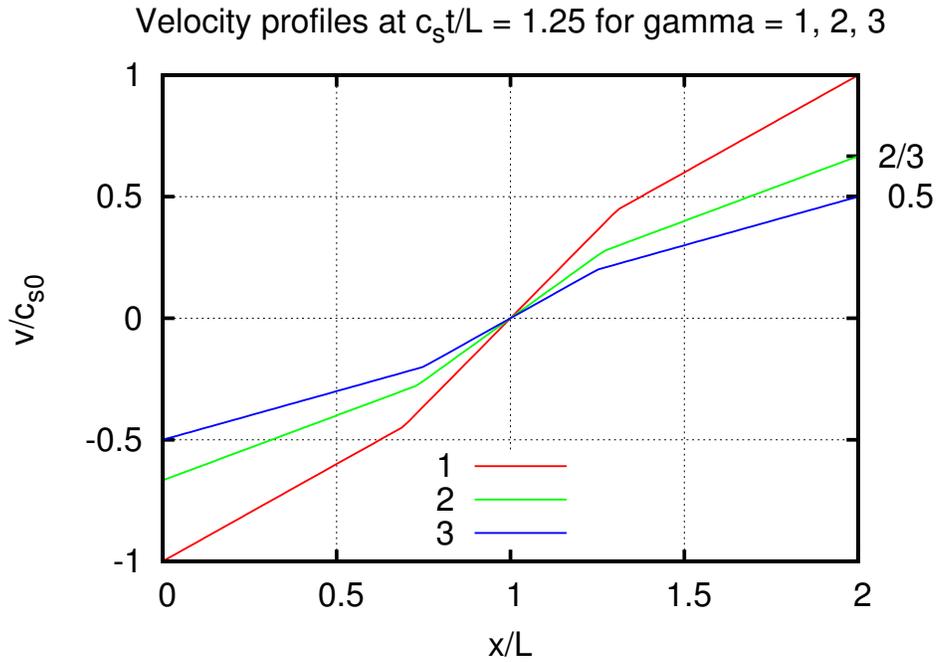}
    \caption{Velocity profiles at $t=1.25L/c_{s0}$ for three values of $\gamma$. Inertial-decay profile is forming in the center.}
    \label{VelAt075}
\end{figure}
\begin{figure}
    \centering
    \includegraphics[scale=0.7, angle=-90]{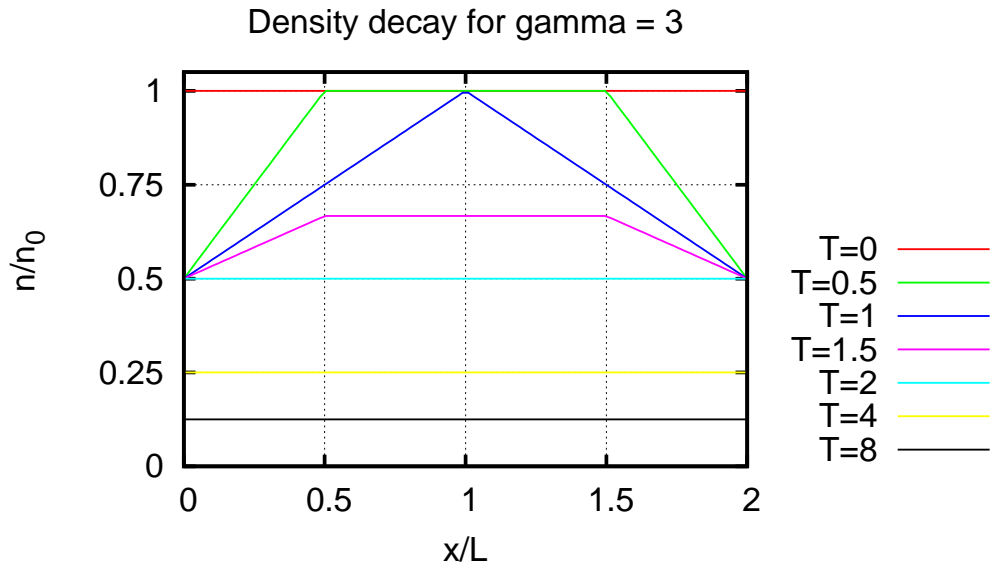}
    \caption{Successive density profiles for $\gamma=3$, exact solution. Flat profile forms at $t=2$ and continues to decay as $1/t$.}
    \label{g3DensEvol}
\end{figure} 
\begin{figure}
    \centering
    \includegraphics[scale=0.7, angle=-90]{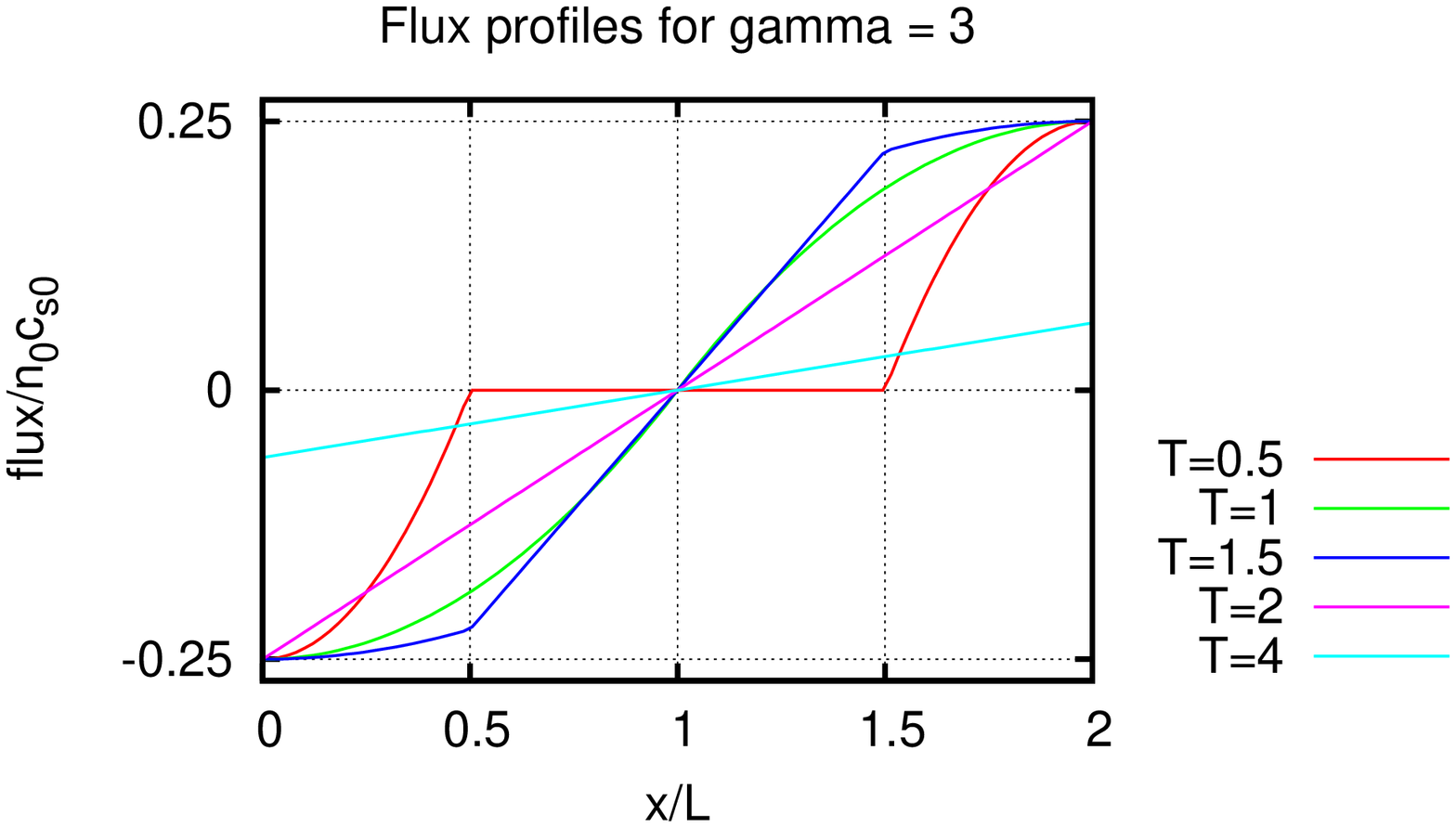}
    \caption{Successive flux profiles for $\gamma=3$, exact solution. Note the stationary point at the boundary for $t<2$. The wall flux decay rate is $1/t^2$.}
    \label{g3fluxes}
\end{figure} 


\begin{acknowledgments}
The information, data, or work presented herein was funded in part by the Advanced Research Projects Agency-Energy (ARPA-E), U.S. Department of Energy, under Award Number DE-AR0001107.
This research was performed at the Princeton Collaborative Low Temperature Plasma Research Facility (PCRF) at PPPL, and supported by the U.S. Department of Energy under contract DE-AC02-09CH11466.
\end{acknowledgments}

\bibliography{Rarefaction}

\end{document}